# Multispectral CT Denoising via Simulation-Trained Deep Learning: Experimental Results at the ESRF BM18

Peter Gänz[1], Steffen Kieß[1], Guangpu Yang, Jajnabalkya Guhathakurta[1], Tanja Pienkny[1], Charls Clark[1], Paul Tafforeau[2], Andreas Balles[3], Astrid Hölzing[3], Simon Zabler[4], Sven Simon[1]

[1] Department of Computational Imaging Systems, ITI, University of Stuttgart, Stuttgart 70569, Germany,
Phone: +49 71168588267, e-mail: peter.gaenz@cis.iti.uni-stuttgart.de, steffen.kiess@cis.iti.uni-stuttgart.de, jajnabalkya.guhathakurta@cis.iti.uni-stuttgart.de, sven.simon@cis.iti.uni-stuttgart.de
[2] European Synchrotron Radiation Facility, Grenoble 38043, France, Phone: +33 438881974, e-mail: paul.tafforeau@esrf.fr
[3] Fraunhofer IIS, division EZRT, Würzburg 97074, Germany, Phone: +49 9313184457, e-mail: andreas.balles@iis.fraunhofer.de, astrid.hoelzing@iis.fraunhofer.de
[4] Imaging Systems / Computed Tomography, Deggendorf Institute of Technology, Deggendorf 94469, Germany, Phone: +49 99136158247, e-mail: simon.zabler@th-deg.de

*Abstract*—Multispectral computed tomography (CT) enables advanced material characterization by acquiring energy-resolved projection data. However, since the incoming X-ray flux is be distributed across multiple narrow energy bins, the photon count per bin is greatly reduced compared to standard energy-integrated imaging. This inevitably introduces substantial noise, which can either prolong acquisition times and make scan durations infeasible or degrade image quality with strong noise artifacts. To address this challenge, we present a dedicated neural network–based denoising approach tailored for multispectral CT projections acquired at the BM18 beamline of the ESRF. The method exploits redundancies across angular, spatial, and spectral domains through specialized sub-networks combined via stacked generalization and an attention mechanism. Non-local similarities in the angular-spatial domain are leveraged alongside correlations between adjacent energy bands in the spectral domain, enabling robust noise suppression while preserving fine structural details. Training was performed exclusively on simulated data replicating the physical and noise characteristics of the BM18 setup, with validation conducted on CT scans of custom-designed phantoms containing both high-Z and low-Z materials. The denoised projections and reconstructions demonstrate substantial improvements in image quality compared to classical denoising methods and baseline CNN models. Quantitative evaluations confirm that the proposed method achieves superior performance across a broad spectral range, generalizing effectively to real-world experimental data while significantly reducing noise without compromising structural fidelity.

*Index Terms*— computed tomography (CT), convolutional neural networks, deep learning, denoising, energy resolved imaging, multispectral imaging, neural networks, synchrotron

## I. INTRODUCTION

This paper addresses the denoising of a multispectral CT setup at the BM18 beamline at ESRF. The presented work and results consists of two conference papers [1,2] that were written by the first author of this paper at different times because the simulations [1] were carried out prior to the beam time experiments in order to prepare for them, and the experimental results [2] were obtained after the beam time. The purpose of this paper is to merge the results of both papers as the proposed neural network requires both simulated data for training and experimental data for inference of the network. Thus, the paper is organized as follows. The rest of the introduction is a literal quote from the introduction of [1] with the last 3 paragraphs from [2] including Figure 1. Section II Related Work and III Proposed Multispectral Denoising Network are literal quotes of [1]. In Section IV, Subsection A. Evaluation on Synthetic Data is a literal quote from [1] and Subsection B. Evaluation on Experimental BM18 Data is a literal quote from [2]. Section V Conclusion discusses and outlines the results achieved and provides insights into future work.

Multispectral computed tomography can contribute in two different ways to advanced computed tomography. On the one hand spectral information can be used for material classification. On the other hand, spectral information can be used for the reduction of CT artifacts such as beam hardening, metal artifacts, and scattered radiation artifacts [3][4]. In the multispectral case the total number of incoming photons at the detector is divided into different energy bins such that the number of photons per bin is significantly lower than the total number detected by a typical energy integration detector in standard CT. In addition, the maximum flux of multispectral detectors (usually photon counting detectors based on Medipix or Timepix) is one or two orders of magnitude lower than standard energy-integrating detectors. The limiting factor is the rate at which the single photons are counted and grouped into individual energy bins. This is because a high photon flux can lead to several absorptions in a pixel at the same time, resulting in an ambiguous energy assignment. Since the detected photons are Poisson distributed and the signal noise is proportional to the square root of the expected signal intensity or the number of detected photons, the SNR of the individual energy bins is significantly lower than that of an energy-integrating detector in classical CT. The distribution of the incoming polychromatic beam causes one of the main problems of multispectral CT, namely the small number of photons in narrow energy bins and the resulting reconstructed volume with very high noise. This effect can be countered by changing the X-ray parameters like increasing the beam intensity. However, in the case of a laboratory-based CT, an increase in intensity also leads to an increase in the size of the focal spot due to cooling limitations and thus to a reduction in spatial resolution.

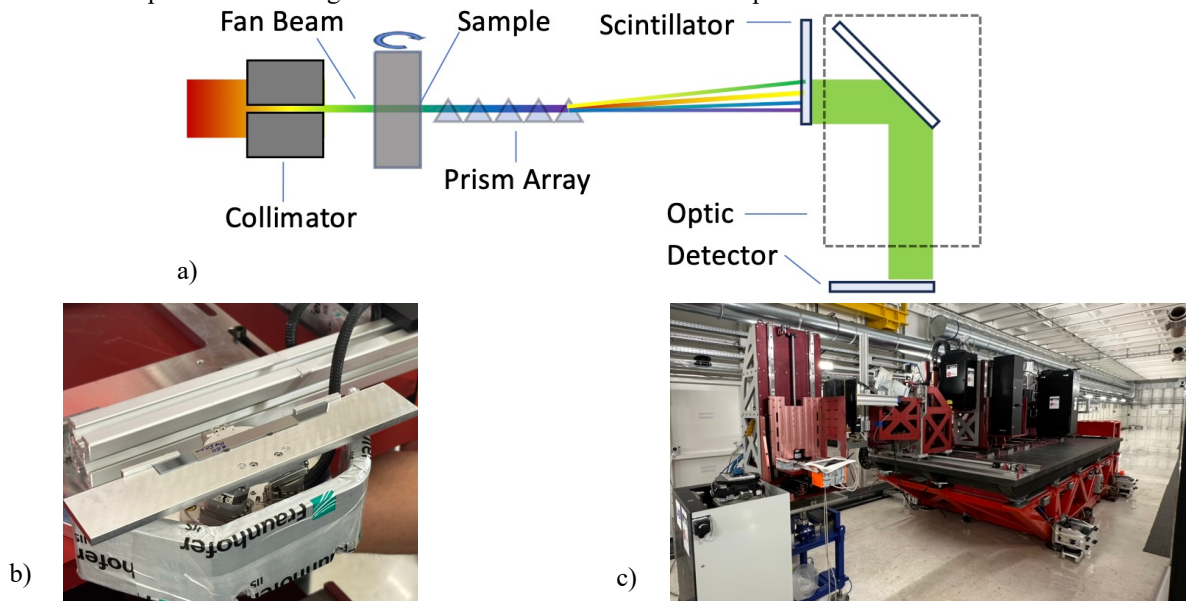

Figure 1 *Multispectral CT setup at the BM18 beamline of the ESRF. (a) Schematic representation of the optical path from the X-ray source to the detector, showing beam collimation, sample interaction, spectral dispersion via the silicon prism array, and detection. (b) Close-up photograph of the mounted silicon prism array on a hexapod stage for precise alignment. (c) Overview of the experimental installation including the sample stage, prism array, and the detector mounted on a girder.*

Alternatively changing exposure time by an order of magnitude or more would extend the scan time from hours to days. Since the impact of such parameter changes is usually unacceptable, we focus on enhancing the SNR by denoising the projection data to enable multispectral CT with an exposure time in the same order of magnitude as with an integrating detector.

The denoising algorithm was developed using the following setup described hereafter, but with modifications it could also be applied to other multispectral CT images, e.g. photon counting detectors in a laboratory CT. The acquisition setup uses a pre-filtered high-energy polychromatic X-ray slice beam with high photon flux at the BM18 beamline of the European Synchrotron Radiation Facility (ESRF) for relatively fast multispectral micro-computed tomography. As shown Figure 1a, the X-ray beam is vertically collimated by high precision slits in order to pass through a $10 - 100$ µm thick slice of the sample. In this setup, the vertical resolution is then linked to the thickness of the slice beam. This is followed by refractive X-ray optics consisting of 50 Si prisms (prism array) behind the object, which vertically spreads the polychromatic beam into its spectral components on a 2D scintillator-based indirect detector. The spectral components are projected onto different horizontal pixel lines of the detector due to the energy-dependent refraction of the prism array. Each detector line, consisting of 2024 pixels with size of $6{,}5$ µm², accumulates a non-equally distributed energy range. In the proposed setup at the BM18 beamline, considering a distance between the prism array and the detector of 10 m and the native spectrum of the beamline, the detector line 0 (the line with the lowest refraction) will contain an energy range from 230 keV to more than 400 keV. Detector lines that are hit by
 strongly refracted beams have smaller energy ranges, e.g. detector line 50 has an energy range from 32.08 keV to 32.4 keV and correspondingly low beam intensities. Figure 2a shows the raw spectrum of the BM18, which hits the sample and then the prism array for refraction. The intensity with neglected air absorption per pixel over the detector lines, which results from the refraction, can be seen in Figure 2b. Here, detector line 0 is irradiated with the highest photon energy of the incident beam, as refraction is low at these energies. Figure 2c shows the average energies across the detector lines and the non-linear distribution of these energies. The low beam intensities lead to very noisy projection images for the low energy ranges and to a poor contrast-to-noise ratio (CNR) for the range with the highest energies. Therefore, noise reduction is required for a CT reconstruction to be used for material decomposition.

We present a method based on neural networks to denoise the projection images, taking into account the similarities of the projection lines of adjacent energy bins as well as the sequential projections.

This network integrates spatial-spectral and spatial-angular neural submodels through an ensemble learning strategy known as stacked generalization, resulting in enhanced denoising capabilities compared to traditional methods and showing improved performance relative to established models such as DnCNN. Although there are notable discrepancies between the simulation setup

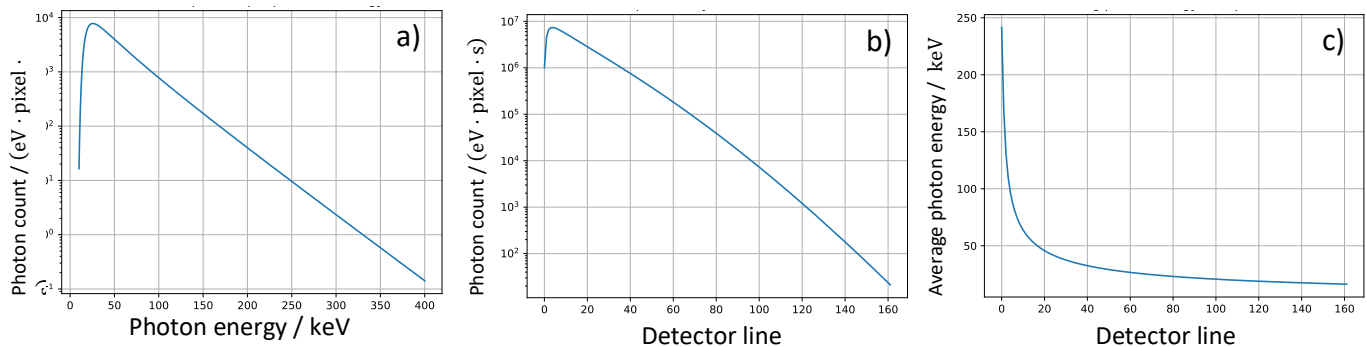

*Figure 2 a) raw BM18 spectrum without prism array or object, b) distribution of the spectrum on the vertical detector lines due to the refraction of the beam by the prism array, c) mean photon energy of each vertical detector line.*

and the actual experimental conditions, the neural network demonstrates reliable generalization and achieves clear improvements in noise reduction, without compromising image detail or introducing additional artifacts.

*A. Multispectral Tomography Experimental Setup at ESRF BM18*

The experimental setup utilized in this study was established at the BM18 beamline of the European Synchrotron Radiation Facility (ESRF). A schematic representation of the setup is shown in a) , while actual photographs of critical components, including the prism array and the detector girder, are presented in Figure 1 b) and c), respectively. Initially, an unfilted polychromatic beam emitted from the BM18 source is horizontaly collimated to achieve a beam height of approximately 20 µm. This precisely collimated slice beam subsequently interacts with the sample positioned on a turntable, enabling sequential angular projections required for computed tomography. Following sample interaction, the attenuated beam subsequently passes through a specially fabricated prism array, which is mounted on a hexapod to allow precise manipulation and alignment. Each prism in this array is fabricated from a (100)-oriented silicon wafer through an etching process, resulting in an angle of 70.52° at the prism tip and side angles of 54.74°. This array, composed of 100 silicon prisms integrated onto a single wafer, refracts the incoming polychromatic beam based on photon energy, effectively dispersing it into distinct energy bins. The cumulative refraction angle, resulting from the serial arrangement of prisms, separates photons by energy levels; lower-energy photons undergo more significant refraction compared to their higher-energy counterparts. An image of the mounted prism array is provided in Figure 1 b). To calibrate the spectral model of the setup, different metal foils were employed. The K-edge energies used for calibration of the refraction model ranged from 26 keV, determined using the K-edge of silver, up to 91 keV, calibrated using the K-edge of bismuth. This calibration procedure of the refreaction model ensures accurate energy assignment across the spectral dimension of the detector. The dispersed beam components are subsequently captured by an sCMOS detector situated 28 meters downstream from the prism array. This detector employs a Gadox scintillator and features an effective pixel size of 14.28 µm. The detector consists of a two-dimensional pixel array, where the vertical direction corresponds to the spectral dimension. Due to the energy-dependent refraction introduced by the prism array, each horizontal pixel line captures a different energy bin. This enables the simultaneous acquisition of multiple energy-resolved projections, although the photon flux varies significantly across the energy bins—higher-energy bins receive substantially more photons than the low-energy ones, which are thus more prone to noise. The mechanical girder that supports the detector, which can be positioned up to 28 meters away from the prism is shown in Figure 1 c). This innovative setup allows us to simultaneously acquire spectral, spatial, and angular data. This data is highly noisy, making the noise suppression methods described in this study necessary.

*B. Challenges in Prism-based Multispectral Imaging*

Despite the advantages of prism-based spectral separation for multispectral imaging, several inherent challenges arise that significantly affect the quality of the acquired data. One major limitation stems from the highly non-uniform photon distribution across energy bins. Due to the energy-dependent refraction introduced by the silicon prism array, lower-energy photons experience greater angular dispersion, leading them to be captured on higher-index detector rows. These detector lines receive substantially fewer photons than those corresponding to high-energy bins, resulting in a dynamic range that spans several orders of magnitude. This disparity and the high absorption of air introduces high noise levels in the lower-energy bins and makes it challenging to achieve satisfactory image quality across the entire spectral domain. Moreover, the spectral resolution of each energy bin is determined by the refraction angle and the vertical pixel size of the detector. While this enables high-resolution energy discrimination in specific spectral regions, the effective width of each bin varies across the detector. This non-linearity in energy bin width complicates direct comparisons between bins and affects data consistency across the detector field.

## II. RELATED WORK

In the following, the setups for the acquisition of multispectral projection data are discussed, as these settings have a significant impact on the energy-dependent intensity, which in turn affects the noise per energy bin. Multispectral setups have been realized both in laboratory-based CT devices and in synchrotron facilities. One of the first spectral CT laboratory devices investigated uses the dual-energy approach. Here, all projections are recorded with 2 different spectra. This can be achieved by different methods, namely by switching the acceleration voltage of the X-ray source so that two complete scans with different spectra are sequentially recorded [4]. Furthermore, the use of two X-ray sources and two detectors, which are offset by 90° from each other, enables the simultaneous recording of the projections of the low and high energy spectra [5]. In addition, the use of two detector layers with different absorption properties enables the simultaneous recording of the two different projection sets with only one X-ray source [6]. The advantage of the dual-energy approach is that the setup is simple and an existing device can be used with software modifications only. The disadvantages are the long scanning time due to successive scans with different acceleration
voltages or the additional need for a second X-ray source and detector. Besides, the two spectra of the projections are very broad and the energy ranges overlap.

With the emergence of photon counting detectors, it became possible to use a conventional X-ray source, e.g. in a laboratory-based CT scanner. Such a detector is able to assign each photon absorbed by a pixel to a specific energy bin or to provide the energy of the photon with single-digit keV accuracy [7][8]. A disadvantage of the photon counting detector is its small size of several cm², high cost and low maximum flux compared to normal integrating detectors. Peng He et al. [9] presented a multispectral micro-CT scanner called Medipix All Resolution System (MARS) with a photon counting detector (Medipix) using eight energy thresholds. They demonstrated a k-edge analysis and a principal component analysis of the reconstructed multi spectral data. The combination of the MARS system with a conventional flat panel detector was used to compensate for the small size of the Medipix detector (14 mm²), see [10]. The flat panel detector is used for a large grayscale overview reconstruction and the photon-counting Medipix detector for a multi spectral region of interest reconstruction.

In synchrotron applications, photon counting detectors can also be used, but due to the low maximum flux of such detectors, other methods are more practical. In [11] a method for reconstructing complete X-ray absorption near-edge structure (XANES) spectra at any position in the sample was presented. For this, the synchrotron beam passes through a fast-switching monochromator used to scan the spectrum, followed by a lens that focuses the beam into the sample to allow local fluorescence. By recording the transmission through a sample and the transmission through the sample and a reference foil with ionization chambers as well as the fluorescence with a positive-intrinsic-negative diode, the local XANES can be reconstructed. Since the beam is focused, it is necessary to scan the beam over the sample in addition to the tomographic rotation. Burza et al. [12] demonstrated the energy dispersion of an X-ray beam using a beryllium prism. The alignment of the beryllium prism to only a part of the X-ray beam enabled a high transmission through a short beam path of the high-intensity center of a Gaussian beam. The experiments showed a high dispersion and a transmission of up to 50%. In [13], a fast tunable monochromator was presented. The monochromator utilizes a prism arrangement in the form of an hourglass to enable a tunable energy range of a beam behind a slit mask. The high rotation speeds of the prism arrangement allow for fast energy scans with a duty cycle of almost 100%.

Image denoising is a subtask of image restoration in computer vision with the objective of removing noise from a noisy image. During acquisition, many different types of noise, such as shot noise, Gaussian noise, compression noise, etc., can degrade image quality. Practical use of CNNs became possible with the GPU-accelerated AlexNet [14], which solved the problems of its predecessor LeNet [15], namely overfitting and high computational cost, by using data augmentation, ReLu activation, and dropout. The presentation of DnCNN [16] demonstrated the superior performance of deep neural networks in denoising compared to the gold standard of non-machine learning based denoising BM3D [17]. The architecture consists of an initial block with a 2D convolutional and ReLu layer, followed by a sequence of 18 hidden layers consisting of 2D convolution, BN, ReLu, and a final convolutional layer to predict a residual noise map. More recently, architectures based on vision transformers such as SwinIR [18] have been proposed, showing equivalent or better results than state of the art CNN based designs such as [19]. Active research is also being conducted in the field of video denoising, where the temporal redundancy between frames is used for denoising, and many promising network architectures have been proposed [20]. DVD-net [21] was the first neural network-based method proposed that exceeds classical block-based algorithms. The model uses neighboring frames and the frame to be denoised from a video sequence and denoises each frame individually. To avoid flicker in the video, the authors introduce a temporal denoising block as a second stage. This second stage was replaced in FastDVDnet [22] by a U-net block, which is able to learn the displacement between the frames. The PaCNet [23] successfully combines the concept of the traditional patch-based method with CNNs. In order to do so, the PaCNet creates artificial frames from a video sequence and uses them to improve its denoising performance. In addition to a spatial denoising block based on convolutional layers, it uses a temporal filter sub model to ensure temporal continuity. The video denoising network PaCNet is relevant for our proposed network, because the tomographic projection sequence is very similar to a video sequence and angular continuity must also be ensured for a CT

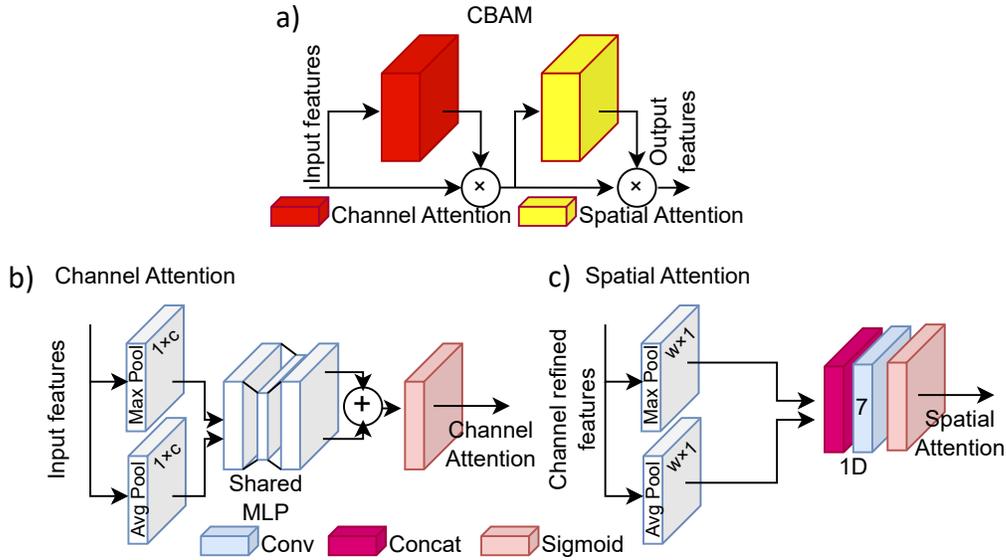

*Figure 3 a) convolutional block attention module overview, b) channel attention squishing the spatial dimension by pooling along the channel direction, c) spatial attention squishing the channel dimension by pooling along the spatial direction.*

reconstruction. Furthermore, neural networks for denoising have also been proposed in the field of hyperspectral imaging (HSI) or remote sensing. The difficulty of fine-tuning the parameters for HSI denoising based on transformations or in the spatial domain was overcome by the proposed HSI-DNet [24]. With an architecture very similar to that of the DnCNN [16], it outperformed the traditional algorithms in terms of speed and accuracy. Similarly, it was proposed by [25] to split the spectral input and process the spectral and spatial features separately by convolution. The processed feature maps are concatenated and fed into 3D convolutional layers with skip connections to learn a residual noise map. Attention-based octave dense network (AODN) [26] was proposed as an attention-based approach. First, the spatial and spectral features are extracted separately by convolutional layers. Then, the concatenated feature maps are successively processed by a channel and a spatial attention module. Multiple octave convolutional blocks are used to split the high and low frequency information before applying a convolution. Another sequence of attention modules and a convolution layer extract the final noise map. The HSI-denoising
network AODN is relevant to our proposed architecture because a multispectral projection also has a spatial and spectral dimension that can be exploited for denoising.

III. PROPOSED MULTISPECTRAL DENOISING NETWORK

The proposed denoising network exploits the spectral, angular and spatial information of a multispectral CT projection dataset.

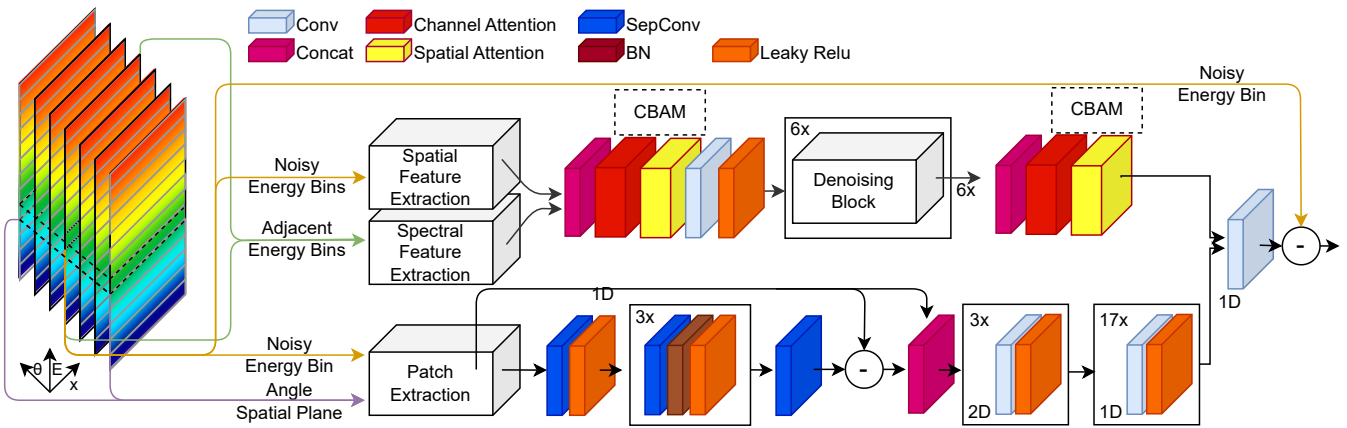

*Figure 4 Denoising model with stacked generalization. The noisy image is denoised by the HSI network and the video denoising network in parallel. The results are combined using a third neural network.*

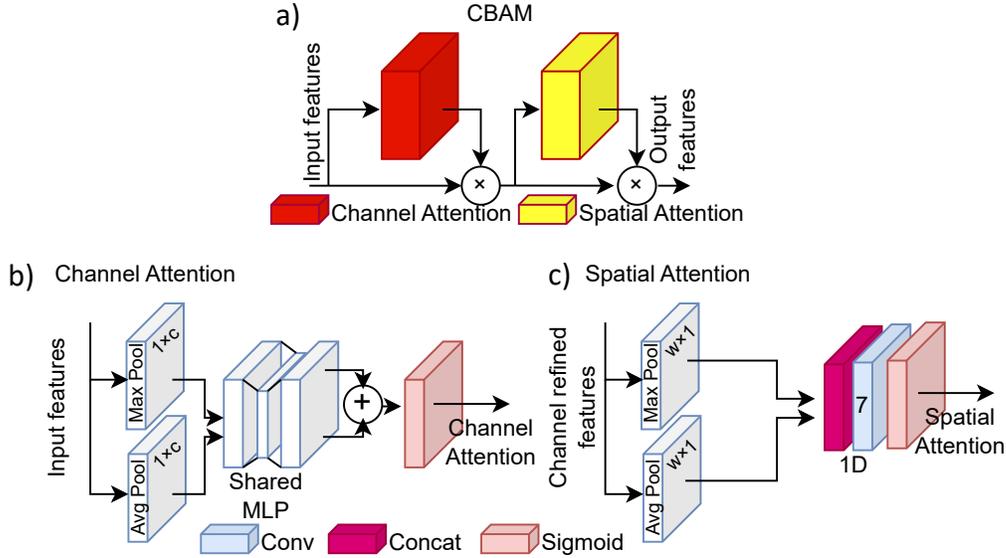

*Figure 5 a) convolutional block attention module overview, b) channel attention squishing the spatial dimension by pooling along the channel direction, c) spatial attention squishing the channel dimension by pooling along the spatial direction.*

In order to reconstruct a volume with reduced noise and without additional artifacts generated by the denoising network, the angular dependencies between the neighboring projections must be taken into account. For example, geometry discontinuities and local or global intensity changes in one of the projections result in reconstruction artifacts. To enable an analysis between the reconstructions of the individual energy bins, the spectral dependencies between neighboring energy bins must also be taken into account. Any discontinuity can be problematic for subsequent material characterization. For these reasons, the neural network architecture, see Figure 3, was combined from two models. The first model uses the complete multispectral projection of a single angular step to denoise an energy band and avoid unexpected value changes between different energy bands. The second model uses adjacent projections of the same energy band to avoid the possible intensity or geometry changes between acquisitions. The models are inspired by the PaCNet [23] designed for denoising video and the AODN [26] model designed for denoising remote sensing or Hyper Spectral Imaging (HSI) data. Due to the structure of our multispectral setup, our raw dataset is a 3D dataset with the shape width × energy × angle for a CT image of a 2D slice. First, the networks are trained independently and afterwards the weights are frozen. The results of the individual networks are combined via stacked ensemble learning [27]. For this purpose, we remove the last layer from both models and use the concatenated predicted feature maps to train the third neural network, which consists of a convolution layer.

*A. Spectral Denoising Network*

Hyperspectral imaging describes the imaging of the same sample in multiple electromagnetic spectra in general in the range from ultraviolet to infrared, where each pixel of a HSI camera captures a spectrum of the object. In the proposed setup, we capture range of the electromagnetic spectrum in the X-ray regime not with one pixel but distributed over multiple vertical detector lines. Due to these similarities the starting point is a state-of-the-art AODN HSI-denoising network [26] which outperforms classical approaches like BM4D [28], LLRT [29], HyRes [30] and FastHyDe [31] as well as deep learning methods like HSIDCNN [32] and ENCAM [33] in HSI restoration tasks in a wide range of noise and artefacts like deadlines. After modification of the architecture, it builds the upper part of the used model as shown in Figure 4. The main components of this sub model are spatial and spectral feature extraction, channel and spatial attention as well as a sequence of denoising blocks. First, spatial features of First, spatial features of one band and projection angle are extracted. The spatial feature extractor consists of 3 1D convolutions in parallel with kernel size 3, 5 and 7 resulting in 3 feature maps, as shown in Figure 6a. Second, spectral features are extracted from $k = 64$ adjacent energy bins. To extract the information from both spectral and spatial domain, two 2D convolutional layer are used. The 2 consecutive 2D convolution layers have a kernel size of $k \times 1$ for the spatial features, followed by a 2D convolution kernel of size $1 \times 3$, $1 \times 5$ and $1 \times 7$ respectively, as shown in Figure 6b.

The extracted feature maps are combined and used as input for the convolutional block attention module (CBAM) [26], [34]. CBAM has been proven to enhance the convolution performance by channel attention, which focuses on the inter-channel relationship, and spatial attention, which focuses on the information location, see Figure 3 [34]. The channel attention module extracts an attention map along the channel by a max pooling and an average pooling layer in parallel, followed by a shared multi-layer perceptron (MLP). Adding the two results and successive squishing by the sigmoid function yields the channel attention map, as shown in Figure 3b. Next the input feature map is pointwise multiplied by the channel attention map and used as input for the subsequent spatial attention module. The spatial attention map is calculated by a max pooling and average pooling layer along the spatial dimension. The output is concatenated and a passed to a vanilla 1D convolution layer with sigmoid activation to generate the spatial attention map, as shown in Figure 3c. Finally, the point wise multiplication yields the CBAM output and is used as input for a chain of denoising blocks. A denoising block contains an Octave Convolution layer followed by a channel and spatial attention layer [26]. The Octave Convolution proposed by [35] reduces the computational and memory cost by dividing the feature channels in a high and low frequency part. The application of vanilla convolution followed by an average pooling operation on a part of the input feature map channels $\alpha \cdot c$ results in low frequency feature maps half the spatial size. The high frequency channel $(1-\alpha) \cdot c$ stay the same size and represent the high frequency feature maps. Multiple of these blocks are chained with skip connections to avoid vanishing gradients. Information exchange from low to high frequency is allowed by a convolution layer followed by a up sampling layer and a from high to low frequency by a convolution layer followed by an average pooling layer. The denoising block from [6] was modified by reducing dimension of the convolution and average pooling to 1D. Further, the final feature map is created by a second CBAM.

*B. Angular Denoising Network*

The second part of the architecture uses the information in different projections for denoising, since only small changes between adjacent projections are expected. This has similarities to the field of video denoising; therefore, our proposed network is inspired by an approach from this field, the PaCNet [23]. It outperforms state of the art classical algorithms like V-BM4F [36] and VNLB [37] as well as deep learning approaches like VNLnet [38], DVDnet [39] and FastDVDnet [40]. A single energy band and multiple projection angles are used as input. The architecture can be seen in the lower part of Figure 4, first patches from all input projections are extracted and rearranged in frames, followed by spatial denoising with multiple convolutional layers whose outputs are subtracted from the extracted patches to enable residual learning. Spatially denoised features are concatenated with the rearranged frames from multiple angular projections in a series of convolutional layers for angular denoising. The patch extraction module extracts all possible overlapping patches from multiple adjacent projections in one energy band and reassembles these patches in new frames. The motivation for the creation of new frames form different patch is to gain different noise patterns for the same projection to enhance denoising [23]. New frames are generated from the n-th nearest neighbor patches which are not overlapping. By the variation of offsets for the starting point of each patch, additional frames can be generated. All created frames, the projection to be processed and (as a reliability metric) the average pointwise square distance between processed projection and frames are concatenated and passed to the denoising network. Due to the large input feature map size the author of [23] proposed a SepConv layer which breaks down one large convolution into 3 sequential convolutions along the spatial, created frame groups and nearest neighbor dimension. First the extracted frames are denoised in the spatial domain by a chain of SepConv, Batch Normalization and ReLU layers, as shown Figure 4. The spatial denoising part is forced to learn the noise by the subtraction of the skip connection. Further the denoised features are passed to a "temporal-post" filter [23], in our setup an angular-post filter to ensure angular continuity. The input of the last part of the network are the previous spatial denoised frames with the angular surrounding frames. The angular denoising network is similar to a DnCNN [16] [23]with a series of 3 2D convolution layers with LeRU activation, a series of 17 1D convolution layers with LeRU activation and a skip connection for residual learning.

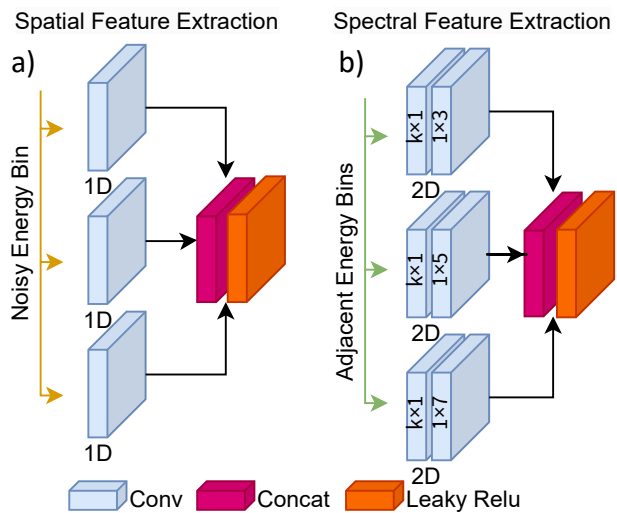

Figure 6 Feature extractor of the spatial denoising path. a) spatial feature extractor with kernel sizes of 3, 5, 7. b) spectral feature extractor first with a kernel size of $k \times 1$ for all three 2D convolutions in the paths followed by $1 \times 3, 1 \times 5, 1 \times 7$ 2D convolutions respectively.

*C. Training Data and Noise Model*

Training the supervised network requires a specific dataset tailored to our setup. The training dataset is created by simulation from 30 simulated slices of three 3D volumes and includes 300 rotation angles, the first 100 detector lines with an energy range of 400 keV to 22.9 keV and 2048 pixel detector width for each slice. In total 9000 projections with 100 energy bins and full detector width are used for training, validation and testing. First the three-3D volume need to be generated. To generate these

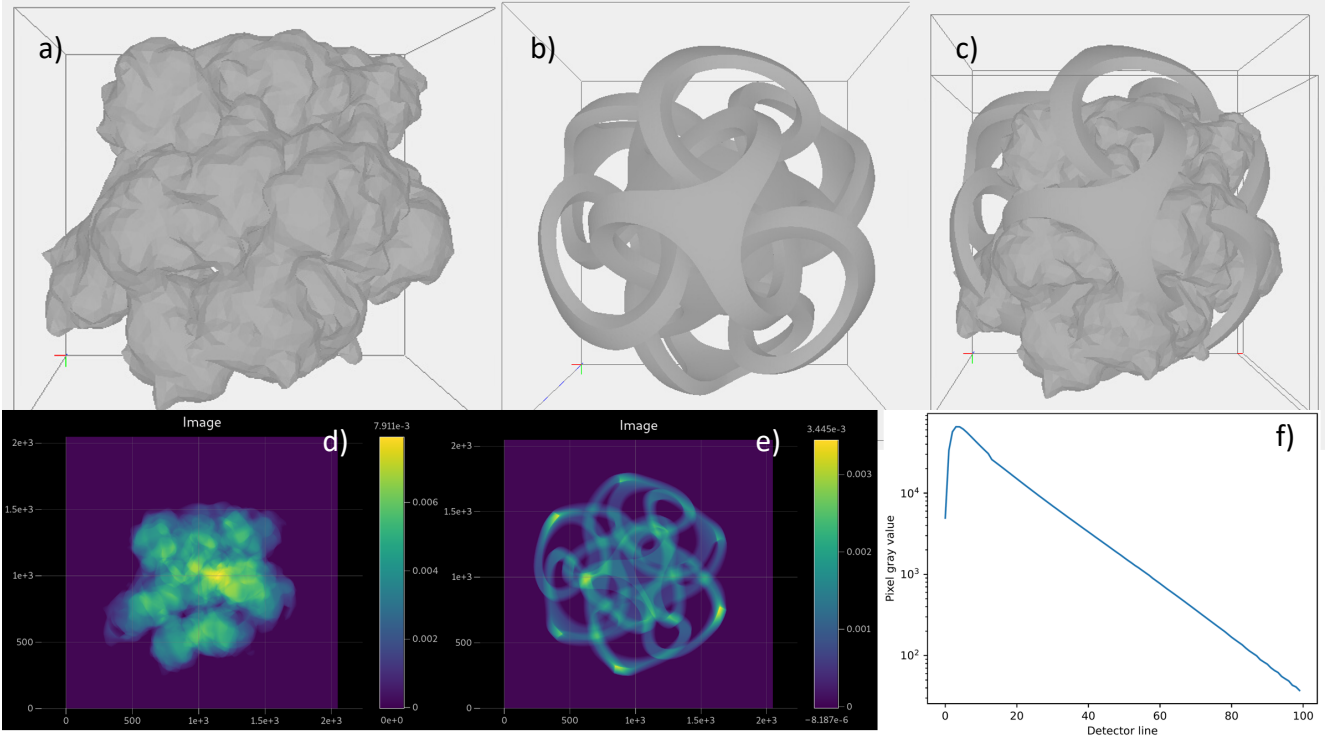

*Figure 7 a) 3D Natural structure, b) 3D industrial structure, c) 3D superimposed structure, d) longitudinal projection of the overlapping free structure which shows the transmission length through the object in m, e) longitudinal projection of the enclosed structure which shows the transmission length through the object in m, f) expected gray value without absorption by a sample on the different detector lines.*

volumes as general as possible, each volume includes natural and industrial structures e.g. a porous stone and an object with sharp edges superimposed, see Figure 7a-c. In order to create such structures, we started with 3D surfaces from Thingi10K [41] and converted them with the inhouse software Voxie [42] to a volume and moved/scaled them to a common bounding box. Next both the overlapping voxel from volume 2 are removed from volume 1 to create an overlapping free volume 1 and an enclosed volume 2. Each volume is then projected with an in-house CT reconstruction software using the setup geometries to a longitudinal projection which represents the transmission length of a ray coming from the source to a certain pixel though the specific volume, resulting in 2D projections as shown in Figure 7e, f. Projections of slices with a distance of 100 voxels were used for further processing.

The longitudinal projection allows a fast conversion to an intensity projection with the BM18 spectrum. Due to the energy dependency of the refraction and absorption the spectrum of the BM18 is processed in 10 eV energy bins. The material of the prism array is Si and the materials in the volumes are Al and SiO2. Next the reduced intensity can be calculated by the Beer-Lambert law. The intensity after the object is described by (1).

$$I_1(E) = I_0(E) \cdot e^{-\mu(E)_1 l_1 - \mu(E)_2 l_2} \quad (1)$$

Here, $I_0$ is the BM18 spectrum at one pixel, $\mu(E)_1$, $\mu(E)_2$ the energy dependent attenuation coefficient of volume 1 and 2 respectively and $l_1$, $l_2$ previous projected transmission length. The absorption by the prism array is calculated shown in (2) and (3).

$$z = 2 \cdot \tan\left(\frac{\alpha_p}{2}\right) \cdot y \cdot n_{pa} \quad (2)$$

$$I_2(E) = I_1 e^{-\mu_{Si}(E) z} \quad (3)$$

$n_{pa} = 50$ is the number of Si prisms with an angle of $\alpha_p = 60°$ and an impinging position of the ray at $y = 0.1$mm. The deflection $\alpha_{pa}$ and corresponding pixel row $Row_{pixel}$ hit by the refracted photons is calculated using (4)-(6).

$$\gamma(E) = \arcsin\left(\sin\frac{\left(\frac{\alpha_p}{2}\right)}{1 - \delta(E)_{Si}}\right) - \frac{\alpha_p}{2} \quad (4)$$

$$\alpha_{pa}(E) \approx \gamma(E) \cdot 2 \cdot n_{pa} \quad (5)$$

$$Row_{pixel}(E) \approx \frac{\alpha_{pa}(E) \cdot d}{h_{pixel}} \quad (6)$$

$\gamma$ is the angle deviation due to refraction on the first prism surface, $\alpha_p$ the angle of the tip of the prism and $\delta_{Si}$ the refractive index decrement of silicon. $\alpha_{pa}$ is an approximation for total refraction angle which is computed by multiplying $\gamma(E)$ with the number of prisms per array $n_{pa}$ times 2 (since two surfaces are impinged). The height of a pixel is described by $h_{pixel}$ and the distance between prism array and detector by $d$. The photons in the same detector row get summed up and if an energy bin overlaps two rows its photon count gets linear distributed, Figure 2c shows the average energy over the detector row. The refraction of the object is neglected. Further, the energy dependent quantum efficiency of conversion of the X-ray photon to optical photons by the LuAG:Ce scintillator is also described by the Beer-Lambert law with the energy dependent attenuation coefficient for $Lu_3Al_5O_{12}$ and a transmission length of 2 mm. The LuAG:Ce scintillator has an energy dependent optical photon yield of $25000 \frac{Photons}{MeV}$. A factor of 4000 as optical loss is assumed before reaching the detector sensor. The intensity peak of the scintillator at 535 nm. The sensor which has its maximum quantum efficiency of 82% at the same wavelength. Finally, released electrons are converted by a factor of $GrayValue = \frac{electrons}{0.46 e/DN}$ [43] to the image gray value for the GT projection. The high intensity at the high energy detector lines due to low refraction and the very low intensity at the low energy range due to high refraction and absorption result in a high dynamic range between the detector lines. Figure 7f shows a profile line of the expected gray value in one pixel with an exposure time of 10 ms in the energy direction of the simulation without object.

Table 1 Expected detector line gray values without object and energy ranges.

| Detector line | Expected Gray value without object | Min. Energy/ keV | Max. Energy / keV |
|---|---|---|---|
| 1st | 4931 | 228.87 | 400 |
| 10th | 41244 | 72.38 | 76,3 |
| 50th | 1603 | 32.4 | 32.73 |
| 100th | 37 | 22.92 | 23.04 |

We have multiple major noise sources in the above-described setup. First, the initial intensity is reduced by the energy dependent absorption of the prisms, the sample and the quantum efficiency of the scintillator. The distribution into separate energy bins and an exposure time of 10ms reduces the intensity further. Poisson noise is applied on the resulting intensity. Next, the scintillator conversion factor yields the intensity of optical photons. Optics reduce the intensity again by a factor of 4000 before applying Poisson noise for a second time. Next, the conversion of the optical photons to electrons by the PCO Edge 4.2 [43] detector is described by its quantum efficiency of 82%. To the resulting number of electrons an offset due to the dark current is added. Further, the photo-response non uniformity is modeled by a gaussian distribution with a standard deviation of 0.001 and an expected value of 1. The dark signal non-uniformity is modeled by a gaussian distribution with standard deviation $\sigma = darkcurrent^2$. The readout noise is modelled with $\sigma = 0.8^2$ [43].

## D. Training

The model is implemented with the open-source library PyTorch. It was trained on 7 Nvidia 1080 Ti GPUs with early stopping after 3 epochs stagnating or increasing of the MSE validation loss. Since we used stacked ensemble learning each described sub model was trained independently. For the training dataset two objects were simulated and for the test data set an additional object was simulated. A hyperparameter search based on Bayesian optimization [44] was used to find the parameters as shown in Table 2. The learning rate (LR) for the stacked ensemble learning was chosen by a factor of 10 smaller than the HSINet LR. To improve the learning over the large range of detector values, all values in a detector line are normalized between 0 and 1. Therefore the calculated maximal possible gray value is used. If this would be done on the whole detector image the high energy detector lines would be close to 1 and the low energy lines close to 0. Due to the significantly smaller values the network would overfit on the high energy data.

Table 2 Hyperparameters

| Hyperparameter | Value |
|---|---|
| Learning rate VideoNet | $1.5 \cdot 10^{-2}$ |
| # Nearest neighbors | 5 |
| Learning rate HSINet | $2.33 \cdot 10^{-5}$ |
| # Adjacent energy bands k | 64 |
| # Denoising blocks | 6 |
| # Octave convolution blocks | 6 |
| Percentage of low frequencies $\alpha$ | 0.1 |

## IV. RESULTS

*A. Evaluation on Synthetic Data*

The trained network was validated with a simulated test dataset with other objects then used for training. The denoised projections show good agreement with the GT projections over a wide energy and dynamic range. For visualization, Figure 8 shows on the left the normalized detector gray value of the noisy projections for the 1st, 10th, 50th and 100th detector line according to our noise model. On the right side the GT and the denoised detector values by our model are detector row and therefore low absorption and low photon count. Figure 8c-h show the denoising results of the network and the good match of the

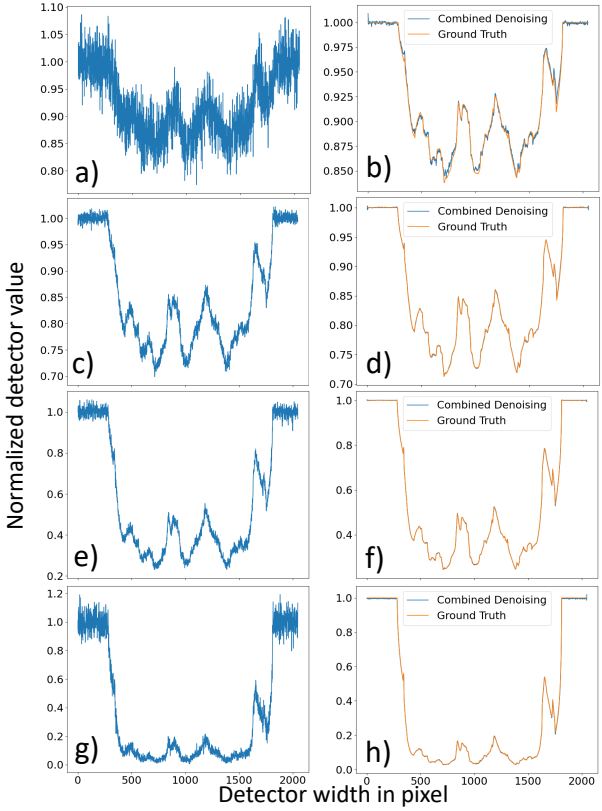

*Figure 8 a) c) e) g) Noisy projections of the 1st, 10th, 50th and 100th energy bin respectively. b), d), f) and h) show the corresponding GT and by our model denoised projections.*

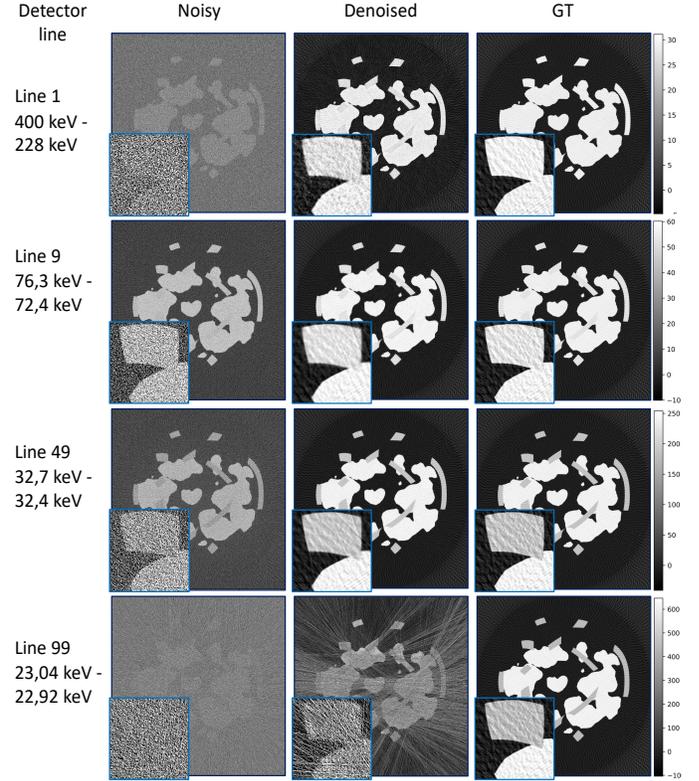

*Figure 9 Reconstruction of the noisy, denoised and GT projections at a single energy band.*

GT data over a wide energy range and high dynamic range. Figure 9 shows the reconstruction of the single energy bin of the 1st, 10th, 50th and 100th detector line for the raw noisy, by our model denoised and GT projections respectively. The reconstruction of the denoised data shows large improvement with no additional artifacts up to detector line 95. A simulated expected gray value of the 16-bit detector without object absorption for the GT projection of line 1st, 10th, 50th and 100th as well as the energy range refracted to this line can be seen Table 1. The limit of the network is reached at an expected gray value without object absorption of about 50 and streak artefacts become visible in the reconstruction. An example of the artefacts is shown in Figure 9 with an expected gray value on detector line 99 of 37. The performance over 100 detector lines were compared with classical methods such as non-local means (NLM) and the total variation (TV) minimization algorithm. Table 3 summarizes the results evaluated by the three classical metrics peak signal to noise ratio (PSNR), structural similarity index measure (SSIM), multi scale SSIM (MS-SSIM)[45] and a deep learning-based metric learned perceptual image patch similarity (LPIPS)[46]. It can be seen that the neural networks individually outperform the classical methods and the combination of the networks produces increases its performance further.

*Table 3 Evaluation of the denoised test set.*

| Evaluation Metric | Noisy  | NLM    | TV     | VideoNet | HSINet | Combined |
|-------------------|--------|--------|--------|----------|--------|----------|
| PSNR              | 36.14  | 38.24  | 45.95  | 49.08    | 54.65  | 54.89    |
| SSIM              | 0.713  | 0.8344 | 0.9764 | 0.9969   | 0.9988 | 0.9986   |
| MS-SSIM           | 0.9151 | 0.9508 | 0.9759 | 0.9984   | 0.9994 | 0.9994   |
| LPIPS             | 0.4127 | 0.2917 | 0.0718 | 0.0099   | 0.0018 | 0.0017   |

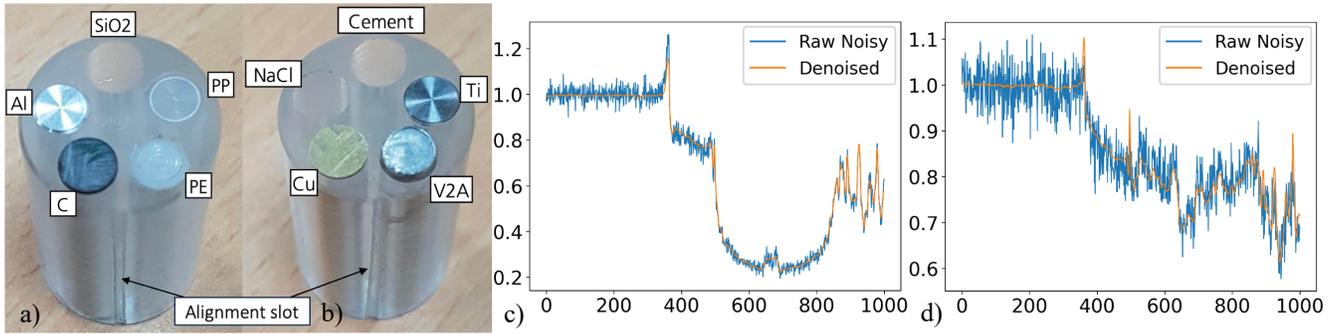

Figure 10. Two cylindrical PP phantoms with five material inserts each, used for validation. a) contains low-Z materials, b) high-Z materials. c) shows the raw and denoised projection of the low Z phantom at energy bin 90., d) the raw and denoised projection at energy bin 0.

B. Evaluation on Experimental BM18 Data

The experimental data were acquired at the BM18 beamline of the ESRF, using two custom-designed phantoms comprising high-Z and low-Z materials [47], [48]. Figure 10 shows photographs of the two phantoms. These phantoms were scanned to generate multispectral datasets, capturing the characteristic attenuation behavior across energy bins ranging from 22.9 keV to 400 keV. Validation was performed using CT scans of phantoms with high and low atomic number (Z) materials:

- High Z: stainless steel V2A, copper, titanium, cement, sodium chloride.
- Low Z: $SiO_2$, polypropylene (PP), polyethylene (PE-HD), aluminum, graphite.

Figure 10 c) and d) illustrate a projection of the low-Z phantom at very low intensity at energy bin 90 (24.02–24.16 keV) and a projection at energy bin 0 with very low contrast-to-noise ratio due to the high photon energy (339–400 keV) respectivly, showing the raw input data with substantial noise and the corresponding denoised result obtained by the proposed model. The suppression of noise while preserving structural features demonstrates the model's effectiveness in low-photon and contrast conditions.

As a comparison, a DnCNN [16] was trained on the same synthetic dataset to denoise individual projections, serving as a single-frame baseline model. Since no ground truth image is available for the real experiment, the experimentally acquired projection data were first reconstructed using filtered backprojection and then evaluated. Performance was quantified by computing the mean normalized RMSE with a 50× averaged reconstruction used as the reference image. Figure 11 displays the raw noisy a), denoised by DnCNN b) and by proposed network c), and reference d) reconstructions for the energy bins 0 of the low-Z phantom.

A visual comparison of the reconstructed images from raw data, 50× averaged reference, and the denoised outputs (proposed model and DnCNN) revealed substantial differences across all evaluated energy bins. Quantitative evaluation was performed using the mean normalized root mean square (NRMSE) with projections 50 times averaged before reconstruction serving as a reference. The proposed network consistently achieved lower RMSE values and exhibited the most effective noise suppression, particularly in low-intensity regions, while maintaining fine structural details, as shown in Table 4. In contrast, the DnCNN showed residual artifacts and streaks, especially at very low signal levels. The raw reconstructions displayed high noise levels, demonstrating the improvements achieved through both learning-based denoising methods, with the proposed network showing clear superiority.

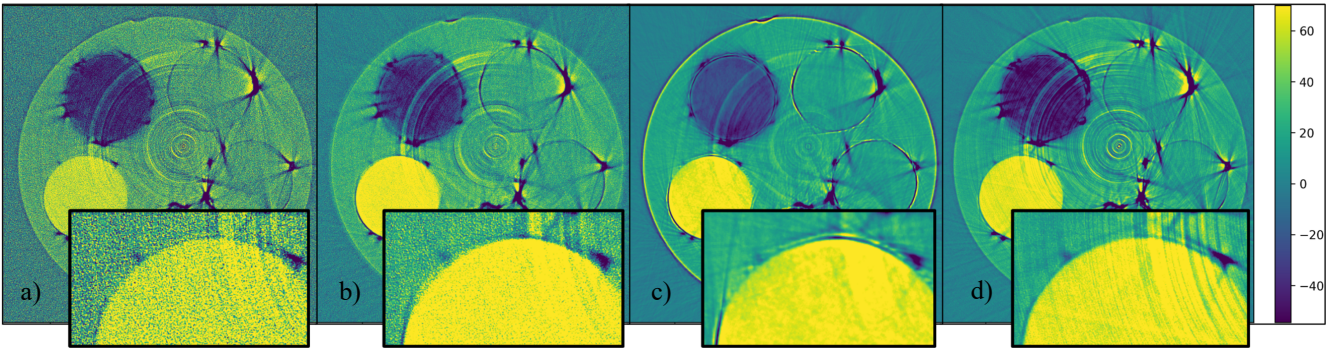

Figure 11 Reconstruction results of the low-Z phantom for energy bin 0. (a) the raw noisy reconstruction, (b) the result after denoising with DnCNN, and (c) the reconstruction using the proposed network and (d) the 50 averaged reference recording.

Table 4 Desnoing performance comparision of the experimental data by NRMSE (smaller is better)

| Phantom | Energy bin 0 | Energy bin 5 | Energy bin 50 | Energy bin 99 |
| --- | --- | --- | --- | --- |
| Low Z Noisy | 5.29 | 2.28 | 1.06 | 1.53 |
| Low Z DnCNN | 3.42 | 2.62 | 0.80 | 1.05 |
| Low Z Proposed | 2.48 | 1.44 | 0.66 | 0.48 |
| High Z Noisy | 3.03 | 1.80 | 6.00 | 3.89 |
| High Z DnCNN | 2.42 | 3.88 | 0.47 | 1.55 |
| High Z Proposed | 0.64 | 0.47 | 0.45 | 0.58 |

Further, the signal-to-noise ratio (SNR) was quantitatively assessed within homogeneous regions of the low-Z phantom. Here, the signal was defined as the mean pixel intensity, and the noise as the standard deviation.

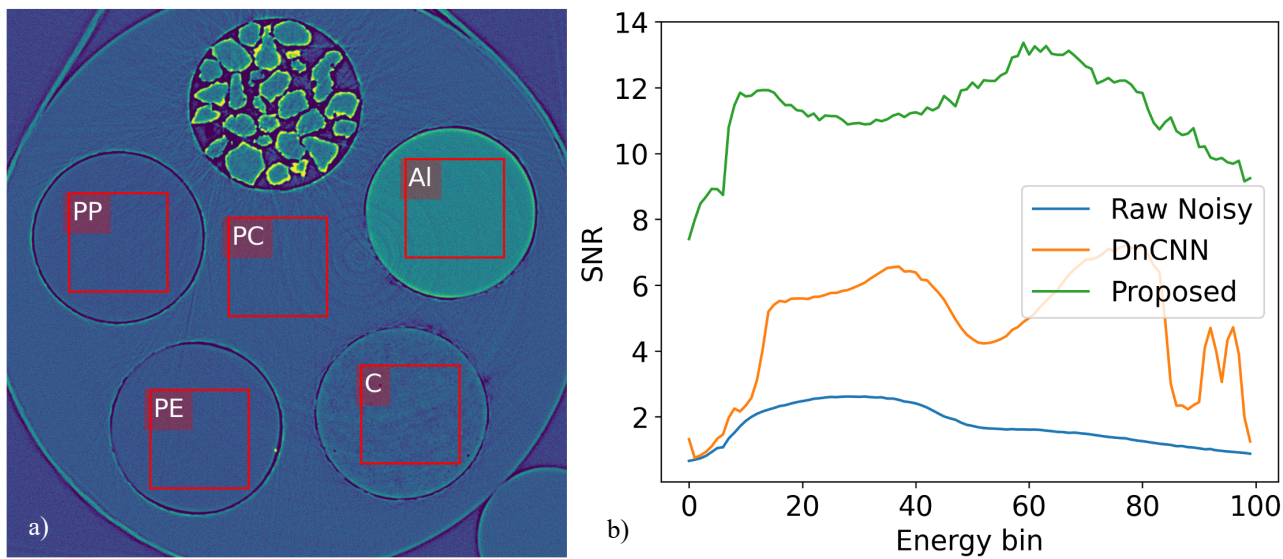

Figure 12 a) Regions of interest (ROIs) used for SNR calculation within the low-Z phantom. b) SNR values over the first 100 energy bins, comparing raw data, DnCNN-denoised, and proposed network-denoised reconstructions.

The SNR was computed separately for homogeneous regions corresponding to different materials in the low-Z phantom across the first 100 energy bins. These per-material SNR values were then combined using root mean square (RMS) to derive a composite SNR metric. Figure 12 b) also displays the SNR for the reconstructed data prior to denoising ("Raw Noisy"). A distinct SNR profile emerges across the energy bins. At low-energy bins, the signal of the low-Z phantom is small due to the combination of high photon energies and low intensity, resulting in low absorption and weak contrast with high noise. The photon flux decreases substantially at high-energy bins due to the narrowing spectral bandwidth, which results in a decrease in intensity and an increase in shot noise. As a result, the SNR is low at both spectral extremes, leading to a characteristic rise and fall of the SNR curve. The results are shown in Figure 12, highlighting the performance of the proposed network compared to both the DnCNN and the noisy reconstructions.

V. CONCLUSION

We presented a neural network architecture for denoising multispectral CT projections acquired under low-photon-flux conditions. By integrating spectral-spatial and angular-spatial submodels via stacked generalization, the method exploits spectral and angular redundancies to deliver effective denoising across a wide energy and dynamic range. Ground-truth training data were synthesized from a public 3D surface database, and realistic noisy inputs were generated with a physics-based noise model of our ESRF BM18 synchrotron setup. Trained on simulations and validated on real BM18 scans of phantoms containing both high-Z and low-Z materials, the approach achieves substantial noise suppression over the full spectral range while preserving structural details; reconstructions exhibit no additional artifacts even for very low amplitudes down to an expected gray value of 0.076% of the detector full scale. Quantitative comparisons show clear improvements over classical denoising algorithms and baseline CNNs, and the observed generalization from simulated to experimental data demonstrates robustness and practical applicability. This advancement enhances the exploitation of low-intensity spectral data for material characterization and represents a step toward high-fidelity multispectral imaging in both synchrotron-based and laboratory CT systems. Future work includes end-to-end fine-tuning by unfreezing pre-trained subnets, increasing angular sampling, rebalancing training data to emphasize low gray values, and extending the framework to denoise 4D spectral data from multispectral CT based on photon-counting detectors.

## Acknowledgements

This work was supported by the Federal Ministry of Education and Research grant number 05K2022, 05K22VSA, 05K22ANA and the State of Baden-Wuerttemberg, Germany. We acknowledge the European Synchrotron Radiation Facility (ESRF) for provision of synchrotron radiation facilities under proposal number MA-6147. This work was partly carried out with the support of the Karlsruhe Nano Micro Facility (KNMFi, www.knmf.kit.edu), a Helmholtz Research Infrastructure at Karlsruhe Institute of Technology (KIT, www.kit.edu).